# Dependable k-coverage algorithms for sensor networks

Gyula Simon[1,3], Miklós Molnár[2], László Gönczy[3], Bernard Cousin[2]

[1] Department of Computer Science, University of Pannonia,
H-8200, Veszprém, Egyetem u. 10, Hungary
[2] IRISA, Campus de Beaulieu, 35 042 Rennes Cedex, France
[3] Dept. of Measurement and Information Systems, Budapest University of Technology and Economics
H-1117 Budapest, XI. Magyar tudósok krt. 2., Hungary
Email: {simon,gonczy}@mit.bme.hu, {molnar, bcousin}@irisa.fr

*Abstract* – *Redundant sensing capabilities are often required in sensor network applications due to various reasons, e.g. robustness, fault tolerance, or increased accuracy. At the same time high sensor redundancy offers the possibility of increasing network lifetime by scheduling sleep intervals for some sensors and still providing continuous service with help of the remaining active sensors. In this paper centralized and distributed algorithms are proposed to solve the k-coverage sensing problem and maximize network lifetime. When physically possible, the proposed robust Controlled Greedy Sleep Algorithm provides guaranteed service independently of node and communication errors in the network. The performance of the algorithm is illustrated and compared to results of a random solution by simulation examples.*

## I. INTRODUCTION

Wireless sensor networks are constructed from small, autonomous sensors and are utilized for various measurement purposes. The individual power capacity of the sensors is very limited (using batteries it may be only some days) but the expected useful lifetime of the network is required to be in the range of weeks or months, depending on the application. To achieve network longevity low duty cycle operation is utilized. In continuously operating sensor networks redundant sensors are deployed from which only a little subset is active at a time; the major part of sensors is turned off and thus saves energy. In redundant dense sensor networks various scheduling algorithms are used to control energy conservation.

In sensor networks used to austerely monitor an area in space it may also be a requirement that *multiple* sensors be able to provide measurements from each point in space. This property may either be necessary because of the applied measurement technology, safety or performance reasons or to satisfy accuracy requirements with relatively low-quality sensors. High redundancy present in the network is necessary to achieve this goal. In general this class of problems can be treated as the k-coverage problem, where coverage means the ability of a sensor to perform measurements over a certain area. While in a dense sensor network there may be several equally good solutions to the *general* k-coverage problem, the energy conservation criterion narrows the range of the acceptable solutions.

In many applications a required coverage is satisfactory as well, so larger coverage does not provide higher performance. Thus finding an 'economical' solution to the k-coverage problem with small number of participating nodes at the same time results energy conservation and thus longer network lifetime as well. Effective scheduling algorithm is required to organize the alternation of active (awake) and sleeping sensor sets to provide continuous service of the network.

In this paper new robust algorithms are proposed to provide dependable k-coverage and prolonged network lifetime. In Section II the main previous results are summarized. Section III introduces a centralized algorithm and its fully distributed variant (Controlled Greedy Sleep Algorithm) to provide robust k-coverage while minimizing the number of awaken sensors at the same time. In Section IV new quality of service metrics are introduced and simulation results are presented to illustrate the capabilities of the proposed algorithms.

## II. PREVIOUS RESULTS

Because of their fragility and power deprivation the dependability of sensor networks is an important and hot research topic. There are several propositions to ensure fault tolerance at different levels [1]. Since the sensors are performing both sensing and communication tasks the main problems of sensor networks are associated to these two activities [14]. The sensor network should be capable of taking measurement in the observed area and transmitting the measured values to sink nodes. The k-coverage problem is associated to the measurement functionality [10]: every point of the target area must be covered by at least *k* sensors (*k* is determined by the application). It also has several implications to connectivity issues [8].

The life-time of the network is generally prolonged by scheduling sleep intervals for some sensors, meanwhile the continuous service is provided by the active sensors (see examples in [5], [6], [7]). The lifetime longevity and the network operability require efficient trade-offs, realized by different scheduling algorithms, which can mainly be divided into two main groups: random and coordinated scheduling algorithms [2]. A distributed, random sleeping algorithm was proposed in [3] where nodes make local decisions on whether to sleep or to join a forwarding backbone, to ensure communications. Each node bases its decision on an estimate of how many of its neighbours will benefit from its being awake and the amount of energy available to it.

In [4] the authors propose a randomized, simple scheduling for dense and mostly sleeping sensor networks. They suppose that there are many redundant sensors in the target area and one can compute the required (identical) duty cycle for individual sensors. In the proposed Randomized Independent Sleeping algorithm, time is divided into periods. At the beginning of each period, each sensor decides whether to go to sleep (with probability $p$ computed from the duty cycle) or not, thus the lifetime of the network is increased by a factor close to $1/p$. This solution is very simple and does not require communication between sensors. The drawback of the proposition is that there is no guarantee for coverage nor for network connectivity. Furthermore, since the sleeping factor is the same for all sensors, this solution cannot adapt to inhomogeneous or mobile sensor setups.

To handle the basic coverage problem the authors in [2] propose a Role-Alternating, Coverage-preserving, Coordinated Sleep Algorithm (RACP). Each sensor sends a message periodically to its neighbourhood containing its location, residual energy and other control information. An explicit acknowledgment-based election algorithm permits to decide the sleep/awake status. The coordinated sleep is more robust and reduces the duty cycle of sensors compared to the random sleep algorithm, and it guarantees 1-coverage in the network. In this solution the topology can affect the behaviour; thus the sensors can adapt their sleeping to the needs. The price of the performance is the significant communication overhead increasing power consumption.

In [9] the asymptotic behavior of coverage in large-scale sensor networks is studied. For the $k$-coverage problem, formulated as a decision problem, polynomial-time algorithms (in terms of the number of sensors) are presented in [10]. A comprehensive study on both coverage and connectivity issues can be found in [11].

In this paper a new coordinated algorithm is proposed that is able to guarantee k-coverage in the network where it is physically possible and at the same time can provide prolonged network lifetime. The proposed algorithm takes into account both the power status and the sensing assignment of the sensors. First, a centralized solution will be proposed, which will be approximated by a distributed algorithm. The latter solution is more feasible in practice due to its low communication overhead.

III. SCHEDULING ALGORITHM

*A. Background*

The sensing assignment in a sensor network can be represented by a bipartite graph $G(R \cup S, E)$, where the two disjoints sets of vertices represent the nodes $S$ and geographical regions $R$ (see Fig. 1). The regions are defined by the subset of sensors that can monitor them. Generally, they cover the whole measured area and are disjoint. In $G$ there is an edge $e$ between region $r \in R$ and sensor $s \in S$ if and only if $s$ (completely) covers region $r$.

The simple *k-coverage problem* is to find a sub-graph $G'(R \cup S', E')$ where $S' \subseteq S$ so that for all vertices $R$ in $G'$ the degree is at least $k$.

The *minimal k-coverage problem* is to find a *non-redundant* sub-graph $G'$ that solves the $k$-coverage problem. A graph $G'$ is non-redundant if there exists no $G''(R \cup S'', E'')$ and $S'' \subset S'$ that solves the simple $k$-coverage problem.

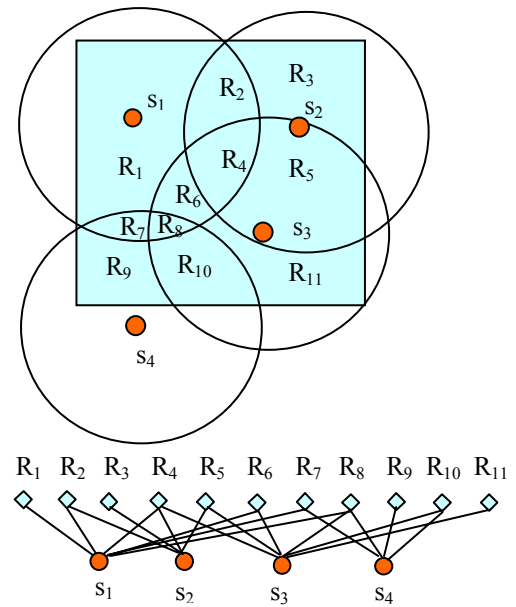

Fig. 1. An example of the target area covered by four sensors ($s_1$, …, $s_4$). The sensing disks and the sensing regions ($R_1$, …, $R_{11}$) are also show, along with the corresponding bipartite graph

*B. Assumptions*

For the sake of simplicity we assume that the coverage of each sensor can be modeled by a sensing disk and a corresponding sensing radius (we use a constant sensing radius but it's not essential). Within its sensing radius a sensor is able to perform measurements, while outside the sensing circle the sensing performance may degrade (but not necessarily).

It's also assumed that the communication radius is at least twice of the sensing radius. In most practical cases it's a sensible assumption and it automatically provides network-wide communication if 1-coverage in sensing is provided [7]. Generally, from the assumption it also follows that network connectivity is higher than $k$ when sensing k-coverage is provided [8].

## C. The centralized k-coverage algorithm

In this version a central scheduler unit is charged with the control of awake/doze states of sensors. The scheduler collects the sensor state information periodically, calculates a special drowsiness factor for each sensor and sends to sleep a subset of sensors, depending on their drowsiness factor. The algorithm is the following:

1. Run the network for a period of $T$
2. Wake up all sensors and collect state information
3. Compute drowsiness factor for each node.
4. Select the node with the largest positive drowsiness factor. Send this node to sleep.
5. Repeat Steps 3-4 while possible (i.e. there is at least one node with positive drowsiness factor).

The drowsiness factor of a node $s$ with current energy $E_s$ is defined as

$$D_s = \begin{cases} \dfrac{1}{E_s^\alpha} \sum_{\forall r \in R} \Phi_r \delta(r,s) & \text{if } \Phi_r > 0, \forall r \\ -1 & \text{otherwise} \end{cases}, \quad (1)$$

where

$$\delta(r,s) = \begin{cases} 1 & \text{if there is an edge in } G \text{ between } s \text{ and } r \\ 0 & \text{otherwise} \end{cases} \quad (2)$$

And $\alpha$ is a positive constant (e.g. $\alpha = 2$), and $\Phi_r$ is the coverage ratio of region $r$ defined as follows:

$$\Phi_r = \begin{cases} \dfrac{1}{c_r - K} & \text{if } c_r > K \\ -1 & \text{otherwise} \end{cases} \quad (3)$$

where $c_r$ is the degree of node $r$ in $G$. The coverage ratio $\Phi_r$ is positive if the region is over-covered, i.e. more than $k$ sensors cover region $r$. $\Phi_r$ is negative if region $r$ is not over-covered: in this case the operation of all sensors covering $r$ is essential.

The drowsiness factor $D_s$ takes into account the energy of sensor $s$: the smaller the energy of a sensor the larger its drowsiness. Negative drowsiness indicates that the sensor is not allowed to sleep.

A sensor participating in many regions that have low over-coverage is likely to participate in more possible solutions than sensors covering regions also covered by many other sensors. Thus a heuristic property is included in $D_s$ to increase the lifetime of the network: sensors participating in regions only slightly over-covered have larger drowsiness. The drowsiness factor for each sensor includes the sum of the coverage ratios of the regions the sensor is able to observe. This property enforces the sensors in critical positions to go to sleep whenever it is possible, to conserve their energy for times when their participation will become inevitable.

Although the centralized algorithm solves the k-coverage problem and conserves the energy of the network at the same time, it assumes network-wide information distribution. In a large network it would mean excessive amount of messages transfer and thus the solution would impose an important communication overhead. Instead, an approximation of the algorithm is proposed that uses information locally available in the neighborhood of a node.

## D. The distributed k-coverage algorithm

The following robust, fault tolerant, distributed algorithm solves the k-coverage problem using locally available information only and thus its communication overhead is low. The algorithm is based on the following observations: To perform approximately the same scheduling as it was shown in the centralized algorithm, a sensor $s$ can go to sleep if its neighbors with larger drowsiness factor decided their state for the next period and $s$ has no critical (not over covered) region to monitor. For this, each sensor should know the drowsiness factor of the neighbors and the decision of neighbors with larger factor. To minimize the local communication, a communication delay (STD) can be associated with each sensor. This delay is inversely proportional with the drowsiness factor. So the sensors with large factor broadcast their decision earlier. Only the awake state decision should be broadcasted explicitly, in this way the communication overhead can be minimal.

*Controlled Greedy Sleep (CGS) Algorithm*
1. Run the network for a period of $T$
2. Wake up all sensors
3. Nodes with energy enough for at least one more period broadcast local Hello messages containing node locations. Based on received Hello messages node $s$ builds up its local set of alive neighbor nodes ($S_s$) with their locations.
4. Each node $s$ calculates its drowsiness factor $D_s$ (see Eq. 1). Instead of the global graph $G$ use the locally known subgraph $G_s(R_s \cup S_s, E_s)$. $R_s$ and $E_s$ are defined in Note 1.
5. Based on $D_s$ each node selects a Shout Time Delay (STD). Small drowsiness means large STD, large drowsiness means small STD.
6. Each node $s$ broadcasts its $STD_s$ and starts collecting other nodes' Awake Messages (AMs). From the received AMs each node builds a List of Awake Nodes (LAN).
7. After $STD_s$ each node $s$ makes a decision based upon the received AMs:
   − if the k-coverage problem can be solved using only nodes present in the LAN and nodes with STD larger than $STD_s$ then *go to sleep*
   − otherwise *stay awake* and broadcast an AM to inform other nodes.

Note 1: For each node the covered regions are represented by a set of squares, as illustrated in Fig. 2. (In the example there are 24 regions, which can be increased to provide better approximation). Let's denote the node's location by $(x_s, y_s)$, and the center of region $i$ by $(\Delta x_i, \Delta y_i)$ in the node's local coordinate system. As a pessimistic approximation, in $G_s$ there is an edge between node $w$ placed at location $(x_w, y_w)$ and region $i$ if

$$R - \sqrt{2}\Delta > \sqrt{(x_s + \Delta x_i - x_w)^2 + (y_s + \Delta y_i - y_w)^2} \quad (4)$$

The square model may seem a rude approximation of the sensing disk, but since the sensing disk model is inherently a rather imperfect estimation of the real sensing area of a sensor, there is no point to put great effort to accurately approximate it. The proposed solution is pessimistic, i.e. certain areas may be unnecessarily over-covered.

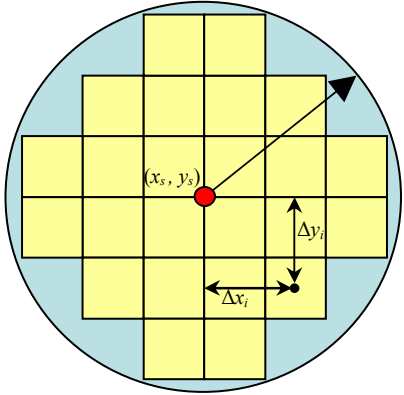

Fig. 2. Possible approximation of the sensing disk of a sensor $s$ by a set of regions $R_s$

Note 2: Each node goes to sleep in a greedy manner: if the coverage problem can be solved with the already known awake neighbors (with higher drowsiness factor) in the LAN (these nodes already have decided on their sleeping status) and some of the neighbors with lower drowsiness factor (these nodes will decide their sleep/awake status later) then the node greedily elects to sleep.

Note 3: The nodes make their decision based upon *received* Hello, STD, and Awake messages. Thus the algorithm is insensitive to lost messages in the sense that the coverage is always provided (if and where it is possible). Naturally, communication problems may cause nodes to stay awake unnecessarily and thus they shorten the lifetime of the network but do not affect the quality of service.

Note 4: The election algorithm tolerates node failures as well. The only situation a failing node can cause problem is when the node dies right after transmitting its STD message. In this case nodes with higher drowsiness factor may incorrectly rely on the presence of the failed node.

Note 5: The communication overhead of the algorithm is low. In each cycle every node broadcasts only at most three messages (two if the node will go to sleep, three otherwise). In addition to this, nodes must stay awake in order to complete the election process. During this extra $T_e$ time nodes consume energy. The communication and awake-time overhead can be neglected if $T$ is significantly longer than $T_e$, which is true in most practical cases.

## IV. RESULTS

### A. Quality of Service Metrics:

The first (functional) requirement is coverage: the network must maintain k-coverage at the largest possible part of the area. The second (non-functional) requirement is the long life-time of the network. The performance of the network can be characterized with the size of the fully covered regions. A possible metric $\Theta_k$ can be the *k-coverage-ratio* defined as

$$\Theta_k = \frac{A_k}{A}, \quad (5)$$

where $A_k$ is the area of the k-covered regions and $A$ is the total area of the target space. If the regions have approximately the same size then another similar metric $\Theta'_k$ can be defined to approximate $\Theta_k$:

$$\Theta'_k = \frac{N_k}{N}, \quad (6)$$

where $N_k$ is the number of the k-covered regions while $N$ is the total number of regions in the target space.

In a critical application the k-coverage must be maintained as long as possible and with $\Theta_k$ as high as possible. If in a degrading network $\Theta_k$ is not satisfactory any more it may still be important to maintain a high value for $\Theta_{k-1}$, $\Theta_{k-2}$, etc, e.g. in order to provide full connectivity in the remaining network.

The *k-lifetime* $L_k(\lambda)$ of a network can be defined as the maximum operational time of the network with $\Theta_k > \lambda$, where $0 < \lambda \leq 1$ (close to 1 in practice).

### B. Simulation results

The proposed CGS algorithm and the random k-coverage algorithm [4] were simulated in Prowler, a probabilistic sensor network simulator [12]. The simulator parameters were set to model the Berkeley MICA motes' MAC layer [13]. The radio propagation model includes realistic effects, e.g. fading, collisions and lost messages.

The tests were performed with a well controlled setup containing 100 nodes placed uniformly on a grid, as can be seen in Fig. 3, showing Prowler's main display. The distance of adjacent nodes on the grid was 10 m, the sensing radius was 15 m, and the communication radius was approximately 40 m (see the parameter settings in Fig. 4). In the simulation the initial energy of all sensors was set to 20 units and in each period awake sensors consumed 1 unit of energy. The period $T$ was set to 1 hour and the required coverage was 3.

In Fig. 5 the performance of the random k-coverage and the CGS algorithms can be compared. The plots show k-coverage ratios $\Theta_k$ for $k = 1,2,3$, as a function of time.

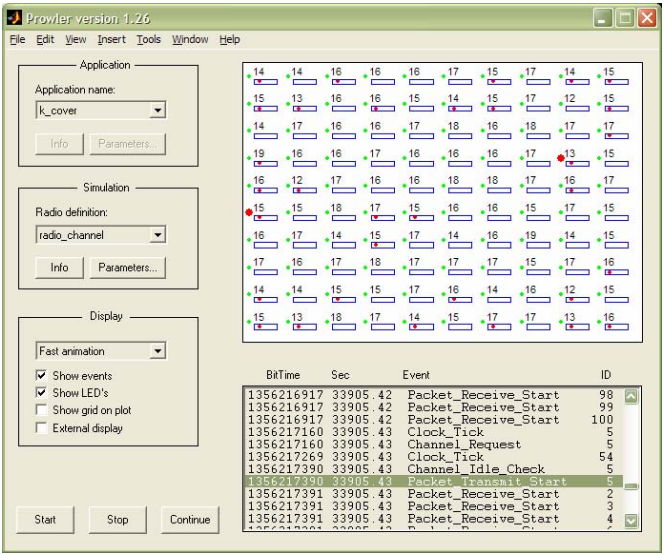

Fig.3. Prowler simulating the 10x10-grid network. Node 5 and 87 (big dots) are transmitting AM messages. Small LEDs in boxes show nodes that have already transmitted AM and will stay awake during the next period. The numbers indicate the actual energy reserve of the nodes.

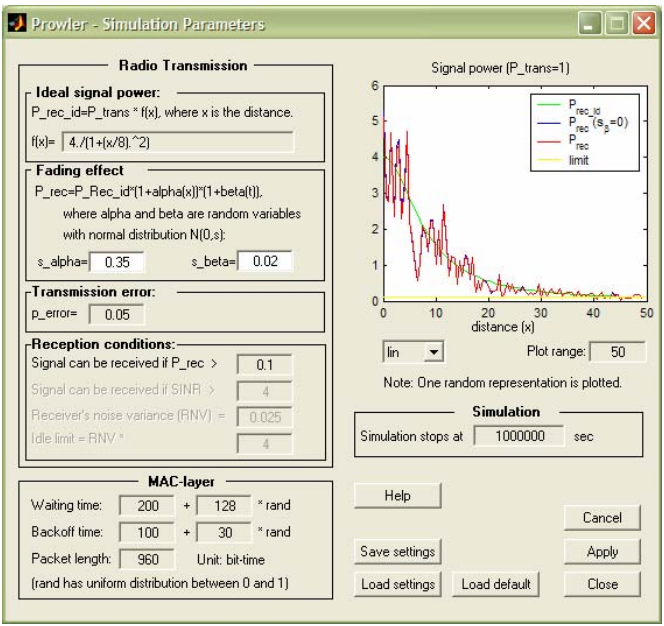

Fig.4. Simulation parameters in Prowler. The plot illustrates a possible (random) signal power vs. distance function

In the experiments the random algorithm's sleeping probability was set to $p_{sleep}$ = 0.4 and $p_{sleep}$ = 0.25. In the first case the lifetime of the random and CGS networks were approximately the same, but while the CGS algorithm managed to provide the required coverage, the random algorithm's 3-coverage ratio was only around 90%. During the service degradation phase (after time instant 26) the CGS still provided much better coverage. When $p_{sleep}$ was set to 0.25 the random algorithm improved its performance (approx. 95%) but the degradation of the network was much more abrupt: at time instant 34 all sensors were completely drained ($\Theta_3 = \Theta_2 = \Theta_1 = 0$), while at this instant for the proposed algorithm $\Theta_3 = 81\%$, $\Theta_2 = 86\%$ and $\Theta_1 = 95\%$.

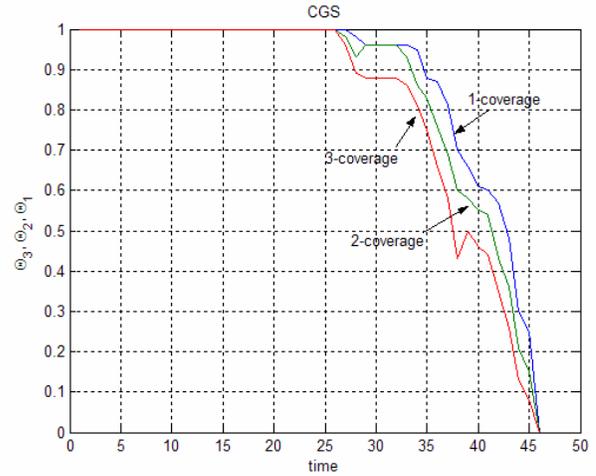

(a)

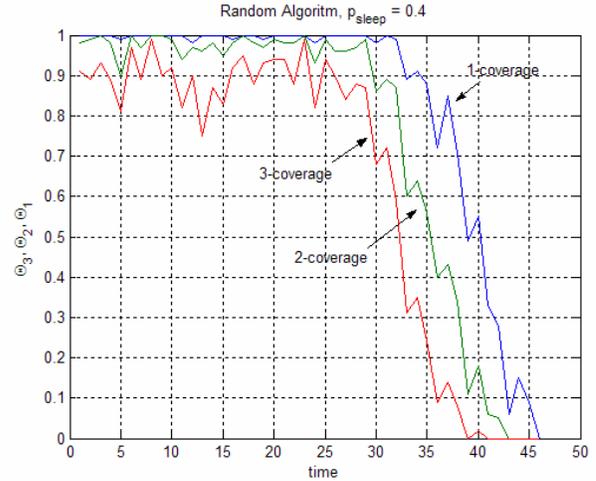

(b)

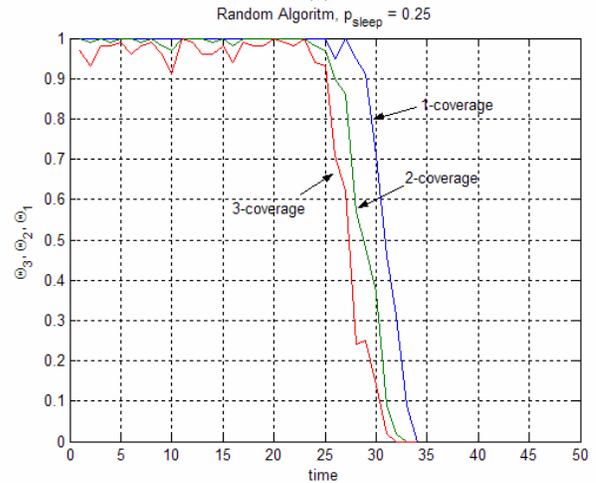

(c)

Fig. 5. Degradation of QoS characteristics of the CGS (a) and the random algorithm with $p_{sleep}$ = 0.4 (b) and $p_{sleep}$ = 0.25 (c), for the sensor network shown in Fig. 3.

The number of awake sensors is shown, as a function of time, in Fig. 6. With $p_{sleep}$ = 0.25 much more sensors were awake in the random network than in the CGS network. To provide similar energy savings similar to the CGS algorithm, the $p_{sleep}$ = 0.4 setting was appropriate for the random algorithm. With this setting, however, the coverage properties were much worse, according to Fig. 5. CGS, however, provides constant good quality service and long network lifetime at the same time.

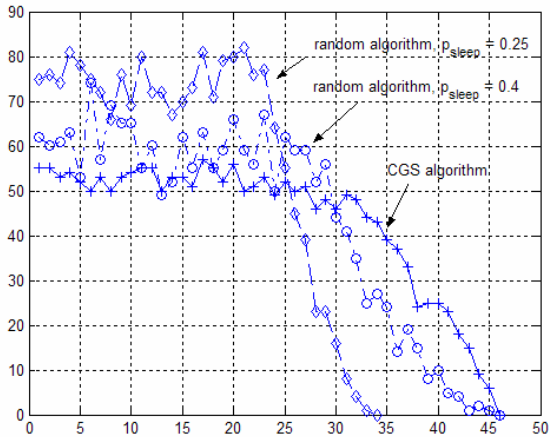

Fig. 6. The number of awake sensors as a function of time for the random and CGS algorithms

## V. SUMMARY

Algorithms were proposed that solve the k-coverage problem and can provide prolonged network lifetime. We showed that the proposed periodic rescheduling of the sleeping and awake nodes saves energy in the network and extends overall network lifetime. The controlled schedule algorithm guarantees k-coverage in the whole network whenever the topology of the network permits it.

The centralized version of the algorithm requires network-wide communication and thus it is not feasible in large networks. A distributed approximation was proposed that uses only locally available information. The Controlled Greedy Sleep Algorithm requires only a few messages to be broadcasted from every node in each period, thus the energy wasted on the communication overhead is small, compared to the gain in the total energy saving in the network, supposing the period of the scheduling is sufficiently large.

The CGS algorithm is robust and fault tolerant: the algorithm provides the required coverage network-wide (if possible) independently of node failures or even high amount of lost messages. The algorithm was compared to the random k-coverage algorithm and was proved to be superior in two senses: while it is possible, the CGS algorithm guarantees the required coverage all over the network. Also, the degradation curve is much gentler, thus the network service is provided for a longer time.

ACKNOWLEDGMENT

This research was partially supported by "Dependable, intelligent networks" (F-30/05) project of the Hungarian-French Intergovernmental S&T Cooperation Program and by the Hungarian Government under contract NKFP2-00018/2005.